\documentclass[prl,aps,twocolumn,notitlepage,superscriptaddress,showpacs,nofootinbib]{revtex4-1}

\usepackage{enumerate,appendix}
\usepackage{amsmath, amsthm, amssymb,commath,braket}
\usepackage{color,calc,graphicx}
\usepackage[usenames,dvipsnames,svgnames,table,cmyk,hyperref]{xcolor}
\usepackage[colorlinks]{hyperref}
\hypersetup{
	colorlinks = true,
	citecolor = {blue},
	urlcolor = {blue},	
	linkcolor = {blue}
}

\usepackage[charter,cal=cmcal,sfscaled=false]{mathdesign}
\usepackage{booktabs}
\usepackage{multirow}
\usepackage{subfigure}
\usepackage{dcolumn}
\usepackage{mathrsfs}
\usepackage{longtable}

\newcommand{\degree}{{}^{\circ}}
\makeatletter

\newcommand{\Rmnum}[1]{\expandafter\@slowromancap\romannumeral #1@}
\makeatother

\begin{document}
\title{Experimental quantum repeater without quantum memory}

\author{Zheng-Da Li}
\affiliation{Shanghai Branch, National Laboratory for Physical Sciences at Microscale and Department of Modern Physics, University of Science and Technology of China, Shanghai 201315, China}
\affiliation{CAS Center for Excellence and Synergetic Innovation Center in Quantum Information and Quantum Physics, University of Science and Technology of China, Shanghai 201315, China}

\author{Rui Zhang}
\affiliation{Shanghai Branch, National Laboratory for Physical Sciences at Microscale and Department of Modern Physics, University of Science and Technology of China, Shanghai 201315, China}
\affiliation{CAS Center for Excellence and Synergetic Innovation Center in Quantum Information and Quantum Physics, University of Science and Technology of China, Shanghai 201315, China}

\author{Xu-Fei Yin}
\affiliation{Shanghai Branch, National Laboratory for Physical Sciences at Microscale and Department of Modern Physics, University of Science and Technology of China, Shanghai 201315, China}
\affiliation{CAS Center for Excellence and Synergetic Innovation Center in Quantum Information and Quantum Physics, University of Science and Technology of China, Shanghai 201315, China}

\author{Yi Hu}
\affiliation{Shanghai Branch, National Laboratory for Physical Sciences at Microscale and Department of Modern Physics, University of Science and Technology of China, Shanghai 201315, China}
\affiliation{CAS Center for Excellence and Synergetic Innovation Center in Quantum Information and Quantum Physics, University of Science and Technology of China, Shanghai 201315, China}

\author{Yu-Qiang Fang}
\affiliation{Shanghai Branch, National Laboratory for Physical Sciences at Microscale and Department of Modern Physics, University of Science and Technology of China, Shanghai 201315, China}
\affiliation{CAS Center for Excellence and Synergetic Innovation Center in Quantum Information and Quantum Physics, University of Science and Technology of China, Shanghai 201315, China}

\author{Yue-Yang Fei}
\affiliation{Shanghai Branch, National Laboratory for Physical Sciences at Microscale and Department of Modern Physics, University of Science and Technology of China, Shanghai 201315, China}
\affiliation{CAS Center for Excellence and Synergetic Innovation Center in Quantum Information and Quantum Physics, University of Science and Technology of China, Shanghai 201315, China}

\author{Xiao Jiang}
\affiliation{Shanghai Branch, National Laboratory for Physical Sciences at Microscale and Department of Modern Physics, University of Science and Technology of China, Shanghai 201315, China}
\affiliation{CAS Center for Excellence and Synergetic Innovation Center in Quantum Information and Quantum Physics, University of Science and Technology of China, Shanghai 201315, China}

\author{Jun Zhang}
\affiliation{Shanghai Branch, National Laboratory for Physical Sciences at Microscale and Department of Modern Physics, University of Science and Technology of China, Shanghai 201315, China}
\affiliation{CAS Center for Excellence and Synergetic Innovation Center in Quantum Information and Quantum Physics, University of Science and Technology of China, Shanghai 201315, China}

\author{Li Li}
\affiliation{Shanghai Branch, National Laboratory for Physical Sciences at Microscale and Department of Modern Physics, University of Science and Technology of China, Shanghai 201315, China}
\affiliation{CAS Center for Excellence and Synergetic Innovation Center in Quantum Information and Quantum Physics, University of Science and Technology of China, Shanghai 201315, China}

\author{Nai-Le Liu}
\affiliation{Shanghai Branch, National Laboratory for Physical Sciences at Microscale and Department of Modern Physics, University of Science and Technology of China, Shanghai 201315, China}
\affiliation{CAS Center for Excellence and Synergetic Innovation Center in Quantum Information and Quantum Physics, University of Science and Technology of China, Shanghai 201315, China}

\author{Feihu Xu}
\email{feihuxu@ustc.edu.cn}
\affiliation{Shanghai Branch, National Laboratory for Physical Sciences at Microscale and Department of Modern Physics, University of Science and Technology of China, Shanghai 201315, China}
\affiliation{CAS Center for Excellence and Synergetic Innovation Center in Quantum Information and Quantum Physics, University of Science and Technology of China, Shanghai 201315, China}

\author{Yu-Ao Chen}
\email{yuaochen@ustc.edu.cn}
\affiliation{Shanghai Branch, National Laboratory for Physical Sciences at Microscale and Department of Modern Physics, University of Science and Technology of China, Shanghai 201315, China}
\affiliation{CAS Center for Excellence and Synergetic Innovation Center in Quantum Information and Quantum Physics, University of Science and Technology of China, Shanghai 201315, China}

\author{Jian-Wei Pan}
\email{pan@ustc.edu.cn}
\affiliation{Shanghai Branch, National Laboratory for Physical Sciences at Microscale and Department of Modern Physics, University of Science and Technology of China, Shanghai 201315, China}
\affiliation{CAS Center for Excellence and Synergetic Innovation Center in Quantum Information and Quantum Physics, University of Science and Technology of China, Shanghai 201315, China}

\begin{abstract}
Quantum repeaters -- important components of a scalable quantum internet -- enable the entanglement to be distributed over long distances. The standard paradigm for a quantum repeater relies on a necessary demanding requirement of quantum memory. Despite significant progress, the limited performance of quantum memory makes practical quantum repeaters still a great challenge. Remarkably, a proposed all-photonic quantum repeater avoids the need for quantum memory by harnessing the graph states in the repeater nodes. Here we perform an experimental demonstration of an all-photonic quantum repeater using linear optics. By manipulating a 12-photon interferometer, we implement a $2\times2$ parallel all-photonic quantum repeater, and observe an 89\% enhancement of entanglement-generation rate over the standard parallel entanglement swapping. These results open a new way towards designing repeaters with efficient single-photon sources and photonic graph states, and suggest that the all-photonic scheme represents an alternative path -- parallel to that of matter-memory-based schemes -- towards realizing practical quantum repeaters.
\end{abstract}

\maketitle

Recent years have seen enormous interest in quantum communication driven by its remarkable features of secure communication~\cite{xu2019quantum}, quantum teleportation~\cite{bouwmeester1997experimental} and distributed quantum computing~\cite{ladd2010quantum}. Photons are considered to be the optimal medium for quantum communication because of their flying nature and compatibility with current telecommunications networks.
However the maximum communication distance is currently severely limited by photon loss in quantum channels, such as optical fibres. One viable solution is to use satellites as relays to transmit photons over a free-space channel~\cite{Yin2017Science,Liao2018PRL}. In fibre-based telecommunications networks, quantum repeaters are believed to be the most promising way to overcome the distance limit~\cite{Gisin2011RMP}. The standard paradigm for a quantum repeater~\cite{BDCZ1998,duan2001} consists of three basic technologies namely entanglement swapping~\cite{Zukowski1993PRL,Pan1998PRL}, entanglement purification~\cite{Pan2001Nature,Pan2003nature} and quantum memory~\cite{chou2007functional,moehring2007entanglement,Yuan2008Nature}. Recently, significant progress has been made both theoretically~\cite{zwerger2012measurement,munro2012quantum,Muralidharan2014PRL} and experimentally~\cite{Chen2017nphoton,Xu2017PRL,kalb2017entanglement}. However, the limited performance of current quantum memories~\cite{Yang2016nphoton} remains a major obstacle in realizing practical quantum repeaters unless there is a future experimental breakthrough.

\begin{figure*}[htbp]
	\centering
	\includegraphics[width=\linewidth]{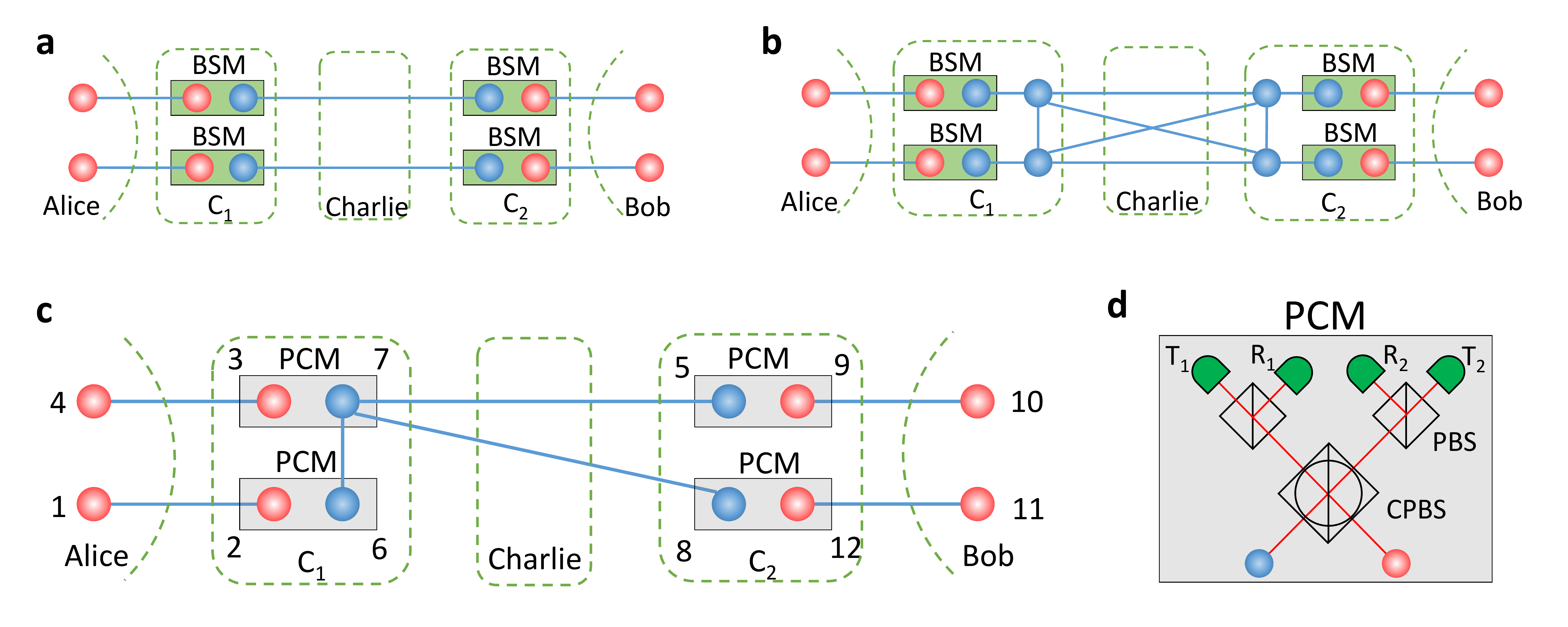}
	\caption{\textbf{Overview of the all-photonic quantum repeater protocol.}
		\textbf{a,} Simple example of conventional quantum repeaters with 2 parallel channels and 1 repeater node (Charlie).
		\textbf{b,} A $2\times2$ example for the all-photonic scheme as proposed in ref.~\cite{azuma2015all} {without loss-tolerant encoding.}
		\textbf{c,} Simplified protocol of a $2\times2$ parallel all-photonic quantum repeater by replacing the RGS with a four-photon GHZ state.
		\textbf{d,} Passive-choice measurement (PCM) device. This performs a Bell state measurement (BSM) when two photons enter the device simultaneously, but a standard projection measurement in $X$ basis with only one photon.
	}
	\label{fig:sscheme}
\end{figure*}

An all-photonic quantum repeater~\cite{azuma2015all} eliminates the need of matter quantum memories. The main concept is analogous to the idea behind measurement-based quantum computation~\cite{raussendorf2001one}. Unlike conventional repeaters (e.g., Fig.~\ref{fig:sscheme}a), the all-photonic scheme introduces an explicit construction of a repeater graph state (RGS) consisting of a complete subgraph of $K$ core photons each connected to an additional photon to form $K$ external arms. Fig.~\ref{fig:sscheme}b shows an example with $K=4$. This approach presents the resilience against photon loss, and also avoids the coherence time limitations of quantum memories and the long-distance heralding requirement. These features have led to all-photonic quantum repeaters attracting much attention recently~\cite{bruschi2014repeat,pant2017rate,Buterakos2017PRX,Ewert2016PRL,Ewert2017PRA,Hasegawa2019ncommun}.

Here, we demonstrate an all-photonic quantum repeater experimentally by manipulating 12 photons generated by 6 independent spontaneous parametric down-conversion (SPDC) crystals. We construct a 12-photon interferometer and verify the ability of manipulating 12 photons by measuring the photon distribution in the $Z$ basis, where $Z$ is the Pauli matrix $\sigma_{Z}$. We successfully demonstrate the concept of all-photonic quantum repeater by realizing an enhancement of entanglement generation rate compared with the conventional parallel entanglement swapping.

First, we explain the theory behind the all-photonic scheme by considering an entanglement swapping protocol, which is an important way of sharing Einstein-Podolsky-Rosen (EPR) pairs over long distances. To obtain higher success rates and longer distribution distances, we can employ $M$ parallel channels and $N$ repeater nodes. Fig.~\ref{fig:sscheme}a shows a simple example with $M=2$ channels and $N=1$ repeater node (Charlie). In such parallel schemes, the entanglement generation rate still decays exponentially with respect to the number of nodes $N$, because a successful event is produced only when all EPR pairs in the same row survive. If each EPR pair has a survival probability $\eta$, the overall entanglement generation rate is $M\eta^{N+1}$. Remarkably, if the repeater nodes can generate an RGS, as shown in Fig.~\ref{fig:sscheme}b, the RGS can serve as a switch that connects EPR pairs in different parallel channels. This enables the entanglement generation rate to reach $M^{N+1}\eta^{N+1}$, which represents an $\mathcal{O}(M^N)$ increase over conventional repeaters without memory. In theory, a large RGS can replace matter quantum memories in a repeater node, thus allowing us to realise an all-photonic quantum repeater~\cite{azuma2015all,pant2017rate,Buterakos2017PRX}.

Demonstrating all-photonic quantum repeater experimentally is challenging, due to the difficulty of preparing a large RGS. To simplify the implementation, we create an experimentally feasible scheme, as shown in Figs.~\ref{fig:sscheme}c and d. In the original proposal (see Fig.~\ref{fig:sscheme}b), the essence of RGS at the repeater nodes is to switch between two functions, (i) establishing entanglement if the entanglement generation succeeds; and (ii) disentangling the qubit if the entanglement generation fails. We utilize the GHZ state and the passive choice measurement (PCM) to realize the switching between those two functions. On the one hand, GHZ state is local-unitary equivalent to a complete graph state. If we perform the Bell state measurement (BSM) between a qubit composing an $m$-partite GHZ state and a qubit composing another $n$-partite GHZ state, we obtain an ($m+n-2$)-partite GHZ state, which establishes the entanglement. If we perform the $X$-basis measurement on the GHZ state, we can disentangle the unwanted qubit. One the other hand, we design the PCM to perform the switching \emph{passively}, in order to make it possible with fewer single photons. Specifically, the PCM performs a BSM when a coincident detection occurs in the two outputs of the circular polarisation beam splitter (CPBS), whereas it automatically performs a standard projection measurement in the $X$ basis when the photon is detected only in one of the two outputs of the CPBS (see Supplementary Information). By doing so, we do not need the active feed-forward operation to decide which photon would be connected or disconnected. Note also that the repeater node, Charlie, uses a delayed preparation of the GHZ-state which means that Charlie prepares the GHZ state just before the arrival of photons from Alice and Bob. This could enable us to assume that the GHZ state is lossless compared with the photons sent from distant nodes of Alice and Bob.

\begin{figure*}[htbp]
	\centering
	\includegraphics[width=\linewidth]{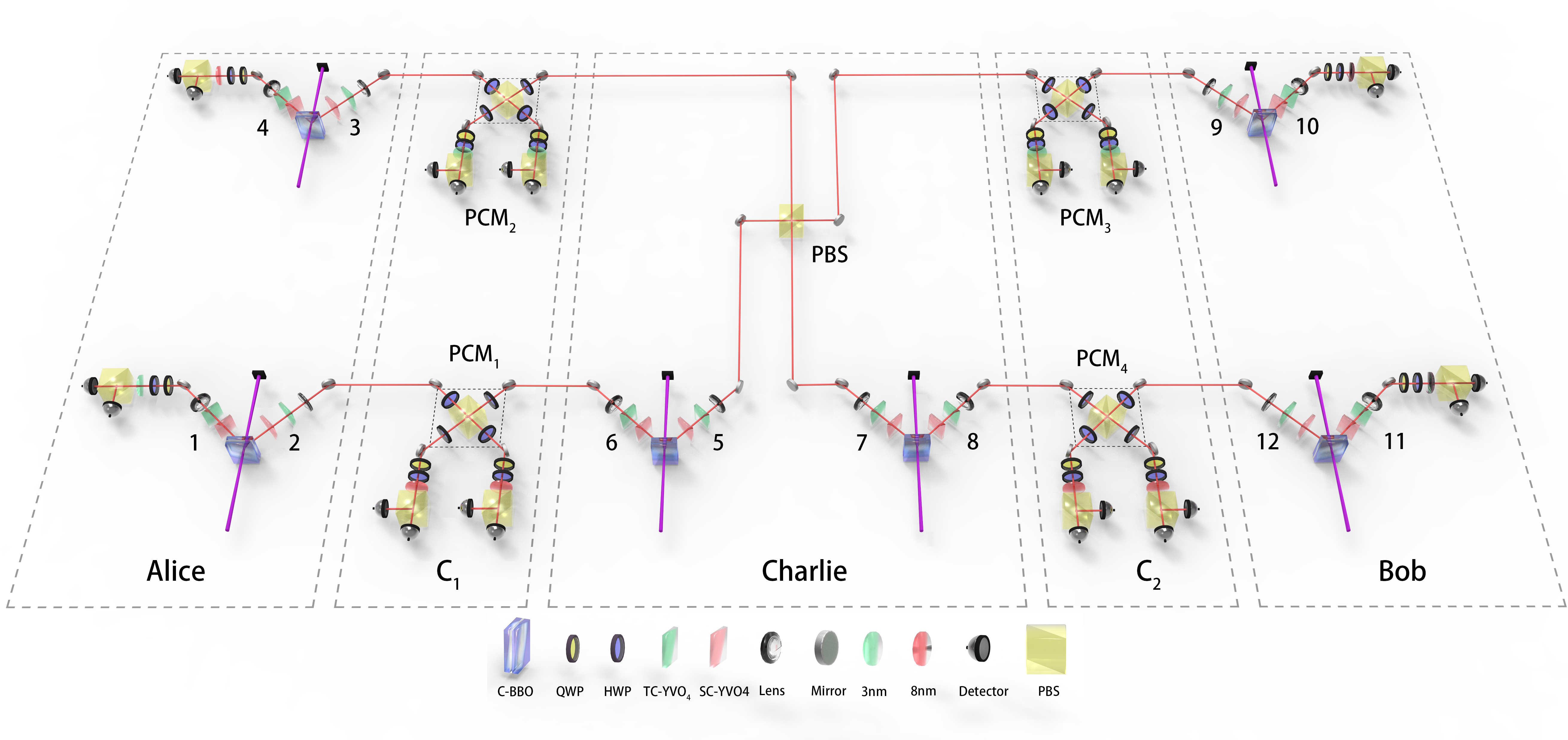}
	\caption{\textbf{The experimental set-up.} An ultrafast ultraviolet (UV) pump laser passes through six separate sandwich-like BBO+HWP+BBO architectures (C-BBO), generating six entangled photon pairs $\ket{\Phi^{+}_{i,j}}=(\ket{HH}+\ket{VV})/\sqrt{2}$ via SPDC processing. Photons 5\&7 are overlapped on a polarising beam splitter (PBS) to prepare a 4-bit GHZ state $\ket{GHZ_4}=(\ket{HHHH}+\ket{VVVV})/\sqrt{2}$. The photons 2\&6, 3\&7, 5\&9, 8\&12 are guided to four PCM devices, each consisting of a circular-PBS (CPBS, implemented by a PBS centered among four HWP at $22.5\degree$), followed by two PBS and four single-photon detectors. Depending on the measurement results , we obtain entangled state between photon pairs 1\&11, 4\&11, 1\&10 and 4\&10. Bandpass filters with FWHM wavelengths of $\lambda_{\text{FWHM}} = 3~\text{nm}$ and $\lambda_{\text{FWHM}} = 8~\text{nm}$ are applied to the e-ray and o-ray photons respectively to remove frequency information. SC-YVO4 and TC-YVO4 represent spatial compensation (SC) and temporal compensation (TC) YVO4 crystals.
	}
	\label{fig:setup}
\end{figure*}

Fig.~\ref{fig:setup} shows an overview of the experimental setup. A pulsed ultraviolet laser with a central wavelength of $390~\text{nm}$, a pulse duration of $150~\text{fs}$ and a repetition rate of $80~\text{MHz}$ is subsequently focused on six sandwich-like combinations of $\beta$-barium borate crystals (C-BBO) to generate six EPR pairs $\ket{\Phi^+_{i,j}}$ via the SPDC processing. Each C-BBO consists of a half wave plate (HWP) sandwiched between two 2-mm-thick, identically cut $\beta$-barium borate crystals. Here, $\ket{\Phi^{+}_{i,j}}=(\ket{HH}+\ket{VV})/\sqrt{2}$, where $\ket{H}$ and $\ket{V}$ denote the horizontal and vertical polarisation states of a single photon. To remove frequency distinguishability among the independent photons, we apply narrow bandpass filters with full width at half maximum (FWHM) wavelengths of $\lambda_{\text{FWHM}} = 3$~nm and $\lambda_{\text{FWHM}} = 8$~nm to the e-ray and o-ray photons respectively. With filtering, the overall system efficiency is quantified to be $38\%$ on average. We typically operate each C-BBO at a down-conversion probability $p= 0.0344\pm0.0001$, obtaining a twofold coincidence counting rate of $3.97\times10^{5}~\text{s}^{-1}$ for each of the 6 EPR pairs with a pump power of $500~\text{mW}$. The fidelity of each EPR pair is above $96\%$.



Of the six EPR pairs, $\ket{\Phi^+_{1,2}}$ and $\ket{\Phi^+_{3,4}}$ ($\ket{\Phi^+_{9,10}}$ and $\ket{\Phi^+_{11,12}}$) belong to Alice (Bob), while $\ket{\Phi^+_{5,6}}$ and $\ket{\Phi^+_{7,8}}$ belong to Charlie. And the EPR pairs of Charlie are used to prepare the four-photon GHZ state. By overlapping the photons 5\&7 on a PBS, we obtain the four-photon GHZ state $\ket{\text{GHZ}_4}=(\ket{HHHH}+\ket{VVVV})/\sqrt{2}$. After preparation, photons 2, 3, 6 and 7 (5, 8, 9 and 12) are send to the node C1 (C2). Then, four PCMs are performed on photons 2\&6, 3\&7, 5\&9, 8\&12. Here, we use movable prisms to adjust the time delays of independent photons.
The CPBS in each PCM device is realised by a normal PBS with four HWPs at $22.5\degree$. A quarter-wave plate, an HWP and a PBS are placed at each CPBS output to perform $Z$ measurements on the photons. Depending on the PCM results, we can obtain final entangled states involving different photon pairs, namely 1\&11, 4\&11, 1\&10 or 4\&10 (see Supplementary Information). So far, the whole setup can be thought of as a 12-photon interferometer. To verify its ability to manipulate 12 photons, we measure the photon distribution in the $Z$ basis and show these results in Supplementary Information.

\begin{figure*}[htbp]
	\centering
	\includegraphics[width=\linewidth]{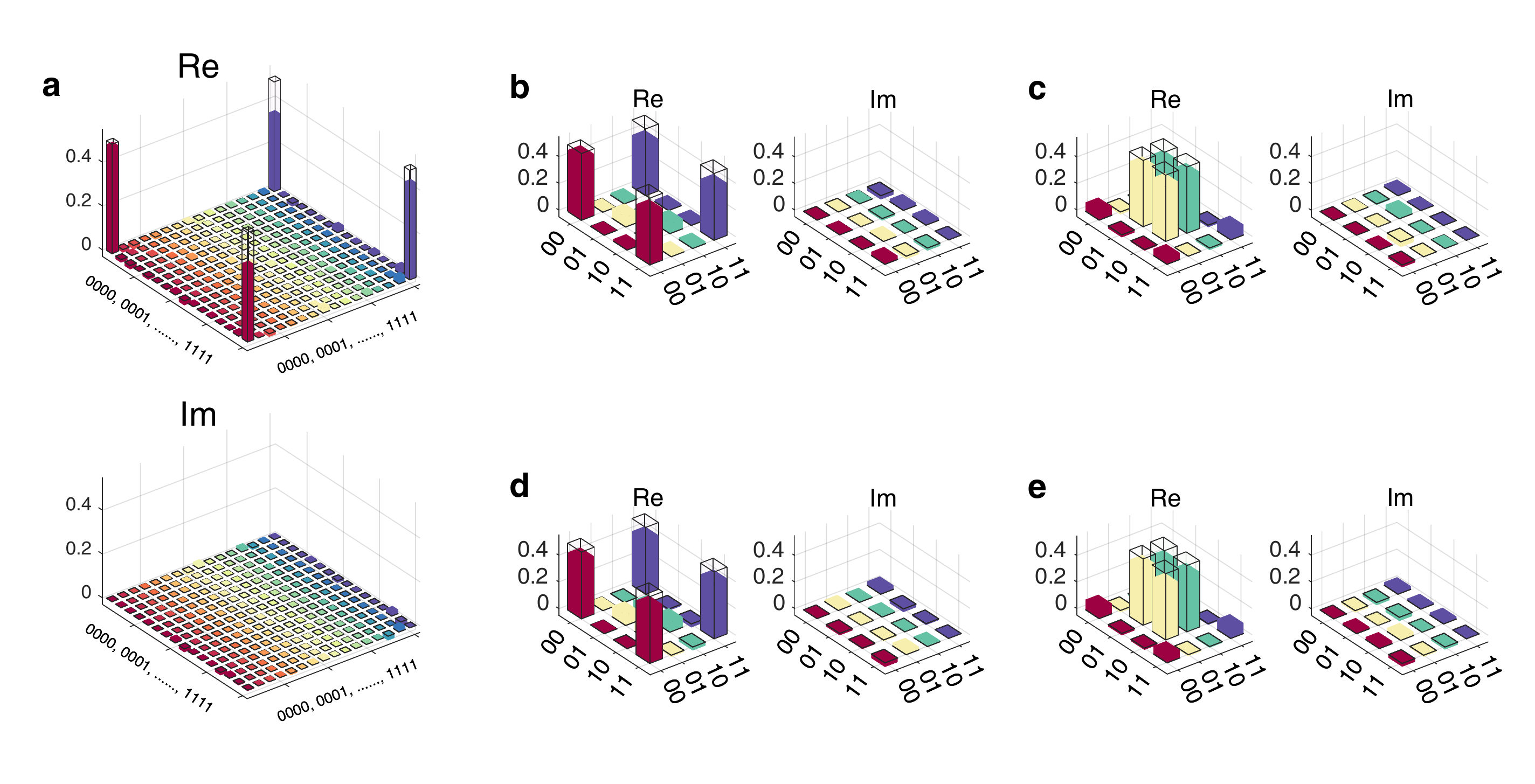}
	\caption{\textbf{Experimental characterisation of the four-photon GHZ state and PCM device.}
		\textbf{a,} Reconstructed density matrix of four-photon GHZ state.
		\textbf{b,c,} Reconstructed BSM operators $\ket{\Phi^+}\bra{\Phi^+}$ and $\ket{\Psi^+}\bra{\Psi^+}$ with $3~\text{nm}$ filters.
		\textbf{d,e,}  Reconstructed BSM operators $\ket{\Phi^+}\bra{\Phi^+}$ and $\ket{\Psi^+}\bra{\Psi^+}$ with $8~\text{nm}$ filters.
		The empty and solid boxes denote the ideal and experimental results, respectively.}
	\label{fig:tomo}
\end{figure*}

\begin{figure*}[htbp]
	\centering
	\includegraphics[width=\linewidth]{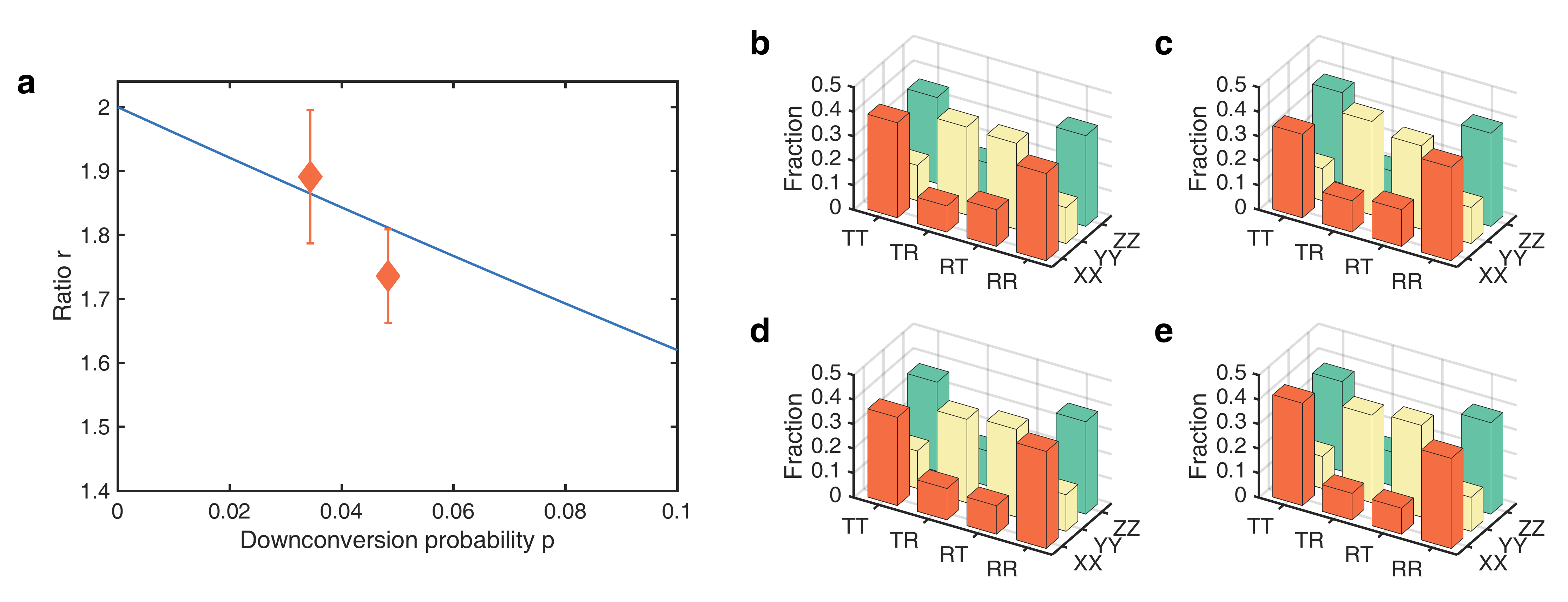}
	\caption{\textbf{Experimental results for the $2\times2$ parallel all-photonic quantum repeater.}
		\textbf{a,} The ratio $r$ of entanglement generation rate between all-photonic quantum repeater and the conventional parallel entanglement swapping, with different down-conversion probabilities $p$. Here, the blue line denotes the theoretical value, whereas the diamonds denote the experiment results: $r=1.89\pm0.10$ for $p=0.0344\pm0.0001$, and $r=1.74\pm0.07$ for $p=0.0483\pm0.0001$. \textbf{b-e,} Measured fractions in $XX$, $YY$ and $ZZ$ bases for entangled photon pairs 1\&11, 4\&11, 1\&10 and 4\&10, respectively. The corresponding fidelities are $0.587\pm0.020$, $0.598\pm0.018$, $0.597\pm0.021$ and $0.628\pm0.019$, respectively.
	}
	\label{fig:result}
\end{figure*}

First, we carry out full tomographic measurements on the four-photon GHZ state to reconstruct the density matrix $\rho_{\text{GHZ}_4}^{\text{re}}$.
Here, we choose the measurement bases $\ket{k}\ket{l}\ket{m}\ket{n}$, where $\ket{k}, \ket{l}, \ket{m}, \ket{n}\in[\ket{H}, \ket{V}, \ket{D}, \ket{A}, \ket{L}, \ket{R}]$, with $\ket{D/A}=1/\sqrt{2}(\ket{H}\pm\ket{V})$ and $\ket{R/L}=1/\sqrt{2}(\ket{H}\pm{i}\ket{V})$. In principle, a total of $1296$ measurement settings are needed. In practice, however, the orthogonal states can be measured simultaneously, meaning that only $81$ measurement settings are required in our experiment. For each setting, we record fourfold coincidences for $60~\text{s}$, yielding a coincidence rate of $370~\text{s}^{-1}$. This enables us to reconstruct the density matrix $\rho_{\text{GHZ}_4}^{\text{re}}$ from the measured data using the maximum likelihood method~\cite{James2001PRA}. The results are shown in Fig.~\ref{fig:tomo}a. Ideally, the density matrix of a four-photon GHZ state would consist of only four real nonzero terms $\ket{0000}\bra{0000}$, $\ket{0000}\bra{1111}$, $\ket{1111}\bra{0000}$ and $\ket{1111}\bra{1111}$, and we can clearly see from Fig.~\ref{fig:tomo}a that the structure of the experimental density matrix is close to the ideal. We also use reconstructed density matrix to calculate the fidelity $F=\bra{GHZ_4}\rho_{\text{GHZ}_4}^{\text{re}}\ket{GHZ_4}=0.896$, which indicates that the prepared four-photon GHZ state is genuinely four-partite entangled.

Next, we characterise the four PCMs experimentally. As is well known that PCM devices can be completely characterised by a measurement operator. If the PCM conduct single-qubit projection measurements, its performance is determined by the extinction ratio of PBS and HWP. In our case, the high extinction ratio of (nearly) $1000:1$ guarantees that the PCM device's projection measurements are correct. When two photons arrive simultaneously at a PCM device, the frequency distinguishability (when not eliminated by filters), is the dominant factor affecting the fidelity. To reconstruct the measurement operators $M_{\Phi^+}^{\text{re}}$ and $M_{\Psi^+}^{\text{re}}$, we perform quantum detector tomography on the PCM~\cite{Luis1999PRLQDT} device, by preparing 16 quantum states $\ket{lm}$, where $\ket{l},\ket{m}\in\{\ket{H},\ket{V},\ket{D},\ket{R}\}$, sending them into the device, and recording the coincidence counts for each state being in $\ket{\Phi^+}$ or $\ket{\Psi^+}$ for $60~\text{s}$. Again, we reconstruct the measurement operators via maximum-likelihood estimation method, and the results are shown in Fig.~\ref{fig:tomo}b-f. For the PCM device with 3 nm filters, the fidelities of $M_{\Phi^+}^{\text{re}}$ and $M_{\Psi^+}^{\text{re}}$ are $F_{\Phi^+}=0.815\pm0.011$ and $F_{\Psi^+}=0.834\pm0.004$, and with 8 nm filters, they are $F_{\Phi^+}=0.819\pm0.015$ and $F_{\Psi^+}=0.813\pm0.015$. The high fidelities of the PCM devices are crucial in implementing all-photonic quantum repeater.

In our experiments, we define the ratio $r$ as entanglement generation rate between the all-photonic scheme and the conventional parallel entanglement swapping. In order to exclude higher-order noise, we just register eight-photon coincidence events. Then, the relationship between $r$ and the down-conversion probability $p$ can be written as $r=2-4p+2p^2$; in the limit as $p$ tends to $0$, $r$ tends to 2 (see Supplementary Information). We record the eightfold coincidence events for the $2\times2$ parallel quantum repeater with $p=0.0344\pm0.0001$ over $39$ hours.
For comparison, we also record these events for the upper (lower) channel of a conventional parallel entanglement swapping by removing CPBS$_2$ and CPBS$_5$ (CPBS$_3$ and CPBS$_4$) with the same $p$ and duration. We evaluate the counting ratio to be $r=1.89\pm0.10$. Then, we increase the power of the pump laser to $720~\text{mW}$ and repeat this experiment with $p=0.0483$ for a duration of $22$ hours, finding a ratio of $r=1.74\pm0.07$. These results are shown in Fig.~\ref{fig:result}a.

Further, to determine the fidelity of entanglement states shared between Alice and Bob, we decompose the density matrix in terms of Pauli matrices:
\begin{equation}
\begin{aligned}
\ket{\Phi^{+}}\bra{\Phi^{+}}=\frac{1}{4}(I+XX-YY+ZZ),\\
\end{aligned}
\end{equation}
where $Z=\ket{H}\bra{H}-\ket{V}\bra{V}$, $X=\ket{D}\bra{D}-\ket{A}\bra{A}$, and $Y=\ket{R}\bra{R}-\ket{L}\bra{L}$.
This means that we only need to measure the state in three bases, i.e., $XX$, $YY$ and $ZZ$. For each bases we record eight-fold coincidence over $39$ hours. The measured fractions for the final entangled photon pairs 1\&11, 4\&11, 1\&10 and 4\&10 are shown in Fig. \ref{fig:result}b-e, respectively. Here, it is important to note that the measured fractions of photon pairs 4\&11, 1\&10 are far away from the uniform distribution, whereas those fractions for the conventional parallel entanglement swapping are uniform without any entanglement. The overall fidelity is $0.606\pm0.010$, which clearly indicates that the final shared state is genuinely entangled. We also measured the fidelity for $p=0.0483\pm0.0001$, finding a value of $0.546\pm0.006$ (see Supplementary Information). Thus, we fully demonstrate a $2\times2$ parallel all-photonic quantum repeater.

To sum up, we have successfully manipulated 12 photons experimentally and accomplished a proof-of-principle demonstration of the all-photonic quantum repeater. Our experiment adopted a GHZ state and a passive scheme to realize the adaptive Bell measurement in the repeater nodes, and achieved an enhancement in entanglement generation rate over the conventional parallel entanglement swapping. Although the all-photonic scheme can remove quantum memories at the intermediate repeater nodes, quantum memories at the end nodes are still needed if Alice and Bob demand a quantum output state. Even so, the memory time at the end nodes required in the all-photonic scheme scales only linearly with communication distance~\cite{azuma2015all}, while the memory time of conventional quantum repeaters scales polynomial or sub-exponential with the communication distance. Note however that, several protocols, such as quantum key distribution~\cite{lo2014secure} and non-local measurements, do not demand strictly a quantum output state, but a shared information. Then, the memories at the end nodes can be removed by using delay-choice entanglement swapping. That is, Alice and Bob measure the local qubits first and wait for the classical signals from intermediate nodes later. In future, the all-photonic scheme is possible to be combined with the matter-memory-based scheme: a RGS can relax the requirement of long coherent time of quantum memory, while a quantum memory can reduce the requirement of large size of RGS. Overall, we believe that all-photonic and matter-memory-based schemes are two important parallel research directions towards a practical quantum repeater.

This work was supported by the National Key Research and Development (R\&D) Plan of China (under Grants No. 2018YFB0504303 and No. 2018YFA0306501), the National Natural Science Foundation of China (under Grants No. 11425417, No. 61771443 and U1738140), Fundamental Research Funds for the Central Universities (WK2340000083), the Anhui Initiative in Quantum Information Technologies and the Chinese Academy of Sciences. The authors particularly thank Hoi-Kwong Lo for insightful discussions and comments.

\onecolumngrid
\newpage

\subsection*{\textbf{\large Supplementary Information for ``Experimental quantum repeater without quantum memory"}}

\section{Memory-based and all-photonic quantum repeaters}

\begin{figure*}[htbp]
	\centering
	\includegraphics[width=\linewidth]{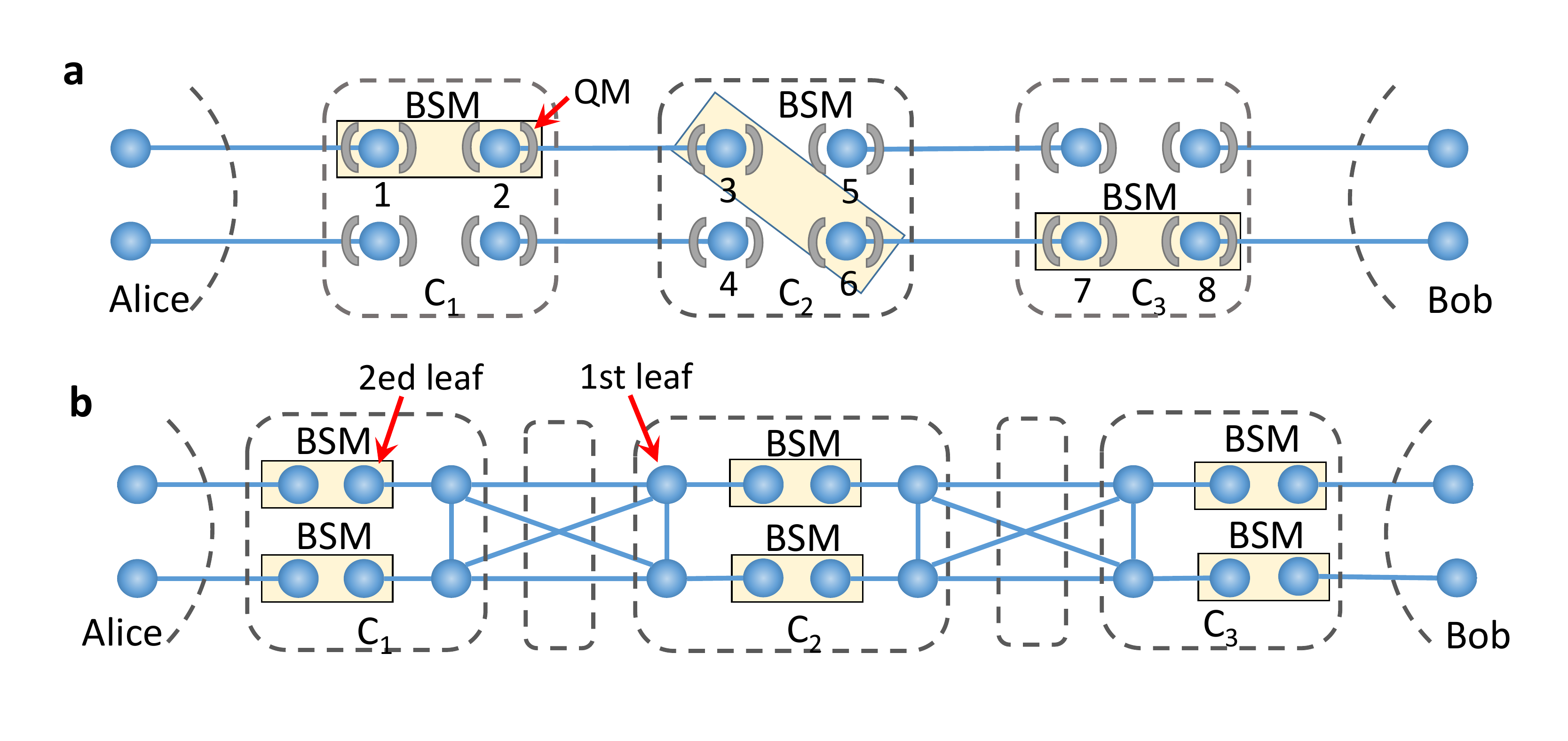}
	\caption{\textbf{The illustration of memory-based and all-photonic quantum repeaters.}
		\textbf{a,} Standard memory-based quantum repeaters. At each intermediate node, the quantum memory (QM) is necessary to keep entanglement until it is informed/heralded of the success/failure of the entanglement generation processes between neighbour nodes. Also, the QM enables the node $C_{i}$ to selectively apply the Bell measurement only to successfully entangled pairs. For instance, QM at $C_{1}$ ($C_{2}$) keeps the photon 1 (photon 8) until it is informed of the success of the entanglement generation from Alice (Bob); QM at $C_{3}$ enables the selective Bell state measurement (BSM) between photon 3 and photon 6, if it is informed of the success of corresponding entanglement swapping from $C_{1}$ and $C_{2}$. \textbf{b,} All-photonic quantum repeaters. At each intermediate node, we use the repeater graph states (RGS) to accomplish the task of selective BSM by a local feed-forward from the 2ed-leaf to 1st-leaf photons. That is, if the BSM succeeds, then we perform $X$ measurement on the corresponding 1st-leaf photon to connect distant photons into the RGS; While the BSM fails, we perform $Z$ measurement on the corresponding 1st-leaf photon to break the 2nd-leaf photon off.}
	\label{fig:comparition}
\end{figure*}

A schematic diagram of the standard memory-based quantum repeaters~\cite{BDCZ1998,duan2001,Gisin2011RMP} is shown in Fig.~\ref{fig:comparition}a. The essence is the quantum memory at the intermediate nodes $C_{i}$, which allows the realization of the \emph{selective} Bell state measurement (BSM) only on qubits that have successfully shared entanglement with distant nodes. Specifically, the quantum memory at $C_{i}$ allows the following two functions. First, the memory enables $C_{i}$ to keep entanglement until it is informed/heralded of the success/failure of the entanglement generation processes between neighbour nodes. Second, the independent accessibility to them enables the node $C_{i}$ to selectively apply the Bell measurement only to successfully entangled pairs.

In the original protocol of all-photonic quantum repeaters~\cite{azuma2015all}, illustrated in Fig.~\ref{fig:comparition}b, one could implement the two functions by using the repeater graph states (RGS) and local feed-forward from the 2nd-leaf to the 1st-leaf qubits. Particularly, if the BSM between the signal photon and the 2nd-leaf photon succeeds, then we perform $X$ measurement on the corresponding 1st-leaf photon in order to connect the distant photons into the RGS. While the BSM fails, we perform $Z$ measurement on the corresponding 1st-leaf photon to break the 2nd-leaf photon off. It is important to note that the local heralding signals are sent and received within the same nodes, which reduces the waiting time to zero, in principle. Therefore, the RGS can replace the quantum memory to realize the selective BSM in a time-reversed manner.

For the end nodes of Alice and Bob, if they demand strictly a shared quantum output state, quantum memories at the end nodes are required in \emph{both} the standard memory-based scheme and the all-photonic scheme to determine which two photons are finally entangled~\cite{azuma2015all}. Note however that in the all-photonic scheme, the heralding signals for connecting different channels are sent and received within the same nodes, so the transmission time of the heralding signals could be nearly zero. Therefore, the memory time at the end nodes in the all-photonic scheme scales only linearly with communication distance~\cite{azuma2015all}. However, the memory time in the standard memory-based scheme scales polynomial or subexponential with the communication distance~\cite{Gisin2011RMP,duan2001}. In our implementation, we use post-selection to determine the final entanglement at the end nodes. Tab.~\ref{Tab:PCMresult} shows the post-selection results.

We remark that several quantum information science protocols, such as quantum key distribution~\cite{lo2014secure} and non-local measurements~\cite{Vaidman2003PRL}, do not demand strictly a quantum output state, but a shared correlated information. Then, the memories at the end nodes can be removed in the all-photonic scheme by introducing the method of delay-choice entanglement swapping. That is, Alice and Bob could measure the local qubits first and wait for the classical signals from intermediate nodes later. This would not lead to redundant errors.

\begin{figure*}[htb]
	\centering
	\includegraphics[width=0.7\linewidth]{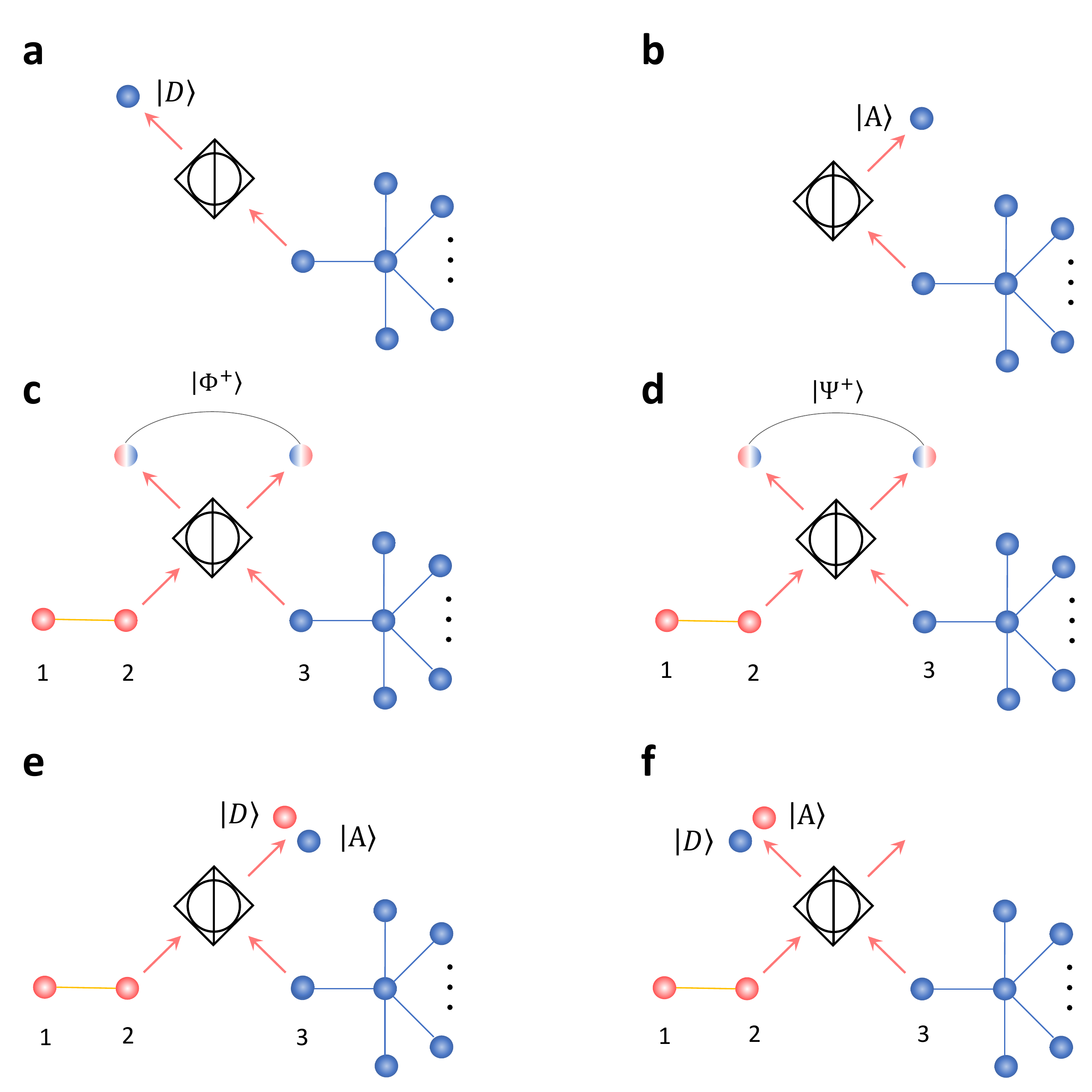}
	\caption{\textbf{The illustration of Circular PBS (CPBS).}}
	\label{fig:CPBS}
\end{figure*}

\section{Passive implementation of the selective BSM}

In the all-photonic scheme, the RGS at the intermediate nodes enables the selective BSM in a time-reversed manner, i.e., switching between two functions: (i) establishing entanglement if the local BSM succeeds; and (ii) disentangling the qubit if the local BSM fails. To simplify the implementation, we create an experimentally feasible scheme. We replace the RGS with the GHZ state, and design the passive-choice-measurement (PCM) to realize the switching \emph{passively}. On the one hand, the GHZ state is local-unitary equivalent to a RGS. If we perform the Bell measurement between a qubit composing an $m$-partite GHZ state and a qubit composing another $n$-partite GHZ state, we obtain an ($m+n-2$)-partite GHZ state, which establishes the entanglement, i.e., function (i). If we perform the $X$-basis measurement on the GHZ state, we can disentangle the unwanted qubit, i.e., function (ii).

One the other hand, we design the PCM to perform the switching between function (i) and function (ii) passively, rather than actively, in order to make it possible with fewer single photons. Specifically, the PCM performs a BSM of function (i), when a coincident detection occurs in the two outputs of the circular polarisation beam splitter (CPBS), whereas it performs a standard projection measurement in the $X$ basis of function (ii), when the photon is detected only in one of the two outputs of the CPBS. We emphasize that when the BSM fails, the PCM device will automatically perform single-bit measurement in $X$ basis. It means that in our passive scheme, we do not need the active feed-forward gate operation to decide which photon would be connected or disconnected. Therefore, we successfully realized the selective BSM in the intermediate nodes, thus demonstrating one of the essences of the all-photonic scheme.

Next, we discuss the implementation of PCM in more detail. We use a circular PBS (CPBS), two normal PBS, and four single photon detectors (SPD) to construct the PCM, as shown in Fig.~\ref{fig:sscheme}d of the main text.

When only one photon from an $N$-qubit GHZ state, and no photon from the EPR pair, is subjected into the CPBS, the GHZ state evolves as
\begin{equation}
\begin{aligned}
\ket{\text{GHZ}^+_N}\bra{\text{GHZ}^+_N}
\xrightarrow{\text{CPBS}}\ket{D^L}\bra{D^L}\otimes\ket{\text{GHZ}^+_{N-1}}\bra{\text{GHZ}^+_{N-1}}+\ket{A^R}\bra{A^R}\otimes\ket{\text{GHZ}^-_{N-1}}\bra{\text{GHZ}^-_{N-1}},
\end{aligned}
\end{equation}
where $\ket{\text{GHZ}^+_N}={1}/{\sqrt{2}}(\ket{H^{\otimes N}}+\ket{V^{\otimes N}})$ and the superscript $L$ ($R$) denotes the photon come out from the left (right) output port of CPBS. In this case, as shown in the Fig.~\ref{fig:CPBS}a-b, the CPBS serves as a single qubit projector in $X$ basis. It is obviously that when the photon is detected only in the right (left) output port, the other photons collapse to a smaller GHZ state $\ket{\text{GHZ}^+_{N-1}}$ ($\ket{\text{GHZ}^-_{N-1}}$) and the entanglement is not destroyed.

When two photon (one comes from the multipartite GHZ state and another one comes from an EPR pair) are subjected into the CPBS, then they evolve as
\begin{equation}
\begin{aligned}
&\ket{\text{EPR}}\bra{\text{EPR}}\otimes\ket{\text{GHZ}^+_N}\bra{\text{GHZ}^+_N}\\
\xrightarrow{\text{CPBS}}&\frac{1}{2}\ket{\text{GHZ}^+_{N}}\bra{\text{GHZ}^+_N}\otimes \ket{\Phi^+_{2,3}}\bra{\Phi^+_{2,3}}
+\frac{1}{2}\ket{\text{GHZ}^+_{N}}\bra{\text{GHZ}^+_N}\otimes\ket{\Psi^+_{2,3}}\bra{\Psi^+_{2,3}}\\
+&\frac{1}{2}\ket{A_{1}}\bra{A_{1}}\otimes \ket{\text{GHZ}^+_{N-1}}\bra{\text{GHZ}^+_{N-1}}\otimes \ket{A^L_{2}D^L_{3}}\bra{A^L_{2}D^L_{3}}\\
+&\frac{1}{2}\ket{D_{1}}\bra{D_{1}}\otimes \ket{\text{GHZ}^-_{N-1}}\bra{\text{GHZ}^-_{N-1}}\otimes \ket{D^R_{2}A^R_{3}}\bra{D^R_{2}A^R_{3}},\\
\end{aligned}
\label{BSM}
\end{equation}
where $\ket{\text{EPR}}=1/\sqrt{2}(\ket{H_{1}H_{2}}+\ket{V_{1}V_{2}})$. The four terms of Eq.~\eqref{BSM} can be distinguished by the CPBS as shown in Fig.~\ref{fig:CPBS}c-f.

In conclusion, we can divide all results above into two cases: (i) When the photons are detected in two output ports, the PCM serves as a bell state analyzer; (ii) When the photons are detected only in left or right output port, the PCM serves as a projector in X basis. It indicates that if the BSM fails, the entanglement of other photons may be not destroyed. The smaller GHZ state $\ket{\text{GHZ}^+_{N-1}}$ or $\ket{\text{GHZ}^-_{N-1}}$ can also be used to connect other EPR pairs.

The analysis above is based on that the detector is perfect (the efficiency is 100\%). However, in a practical apparatus, the efficiency should not be $100\%$. With the imperfect device, we introduce an extra error into the result. In our experiment, the overall system efficiency is 38\% in average and the down-conversion probability is 0.0344 and 0.0483. Then the error rate can be estimated as 1.04\% and 1.45\%. It can be ignored in our experiment.

\section{The corresponding final states with different results of four PCMs}

In our scheme, with different results of PCM, we may obtain final entangled states with different photon pairs, namely 1\&11, 4\&11, 1\&10 or 4\&10. The table of respective final state according to different results of four PCMs is shown in Table~\ref{Tab:PCMresult}.

\setlength{\tabcolsep}{5mm}{
	\begin{longtable}{|c|c|c|c|c|c|c|c||c|c|c|}
		\caption{\textbf{Final states with different PCM results.}} \label{Tab:PCMresult}\\
		\hline
		\multicolumn{2}{|c|}{$\text{PCM}_1$} & \multicolumn{2}{|c|}{$\text{PCM}_2$} & \multicolumn{2}{|c|}{$\text{PCM}_3$} & \multicolumn{2}{|c||}{$\text{PCM}_4$} & \multicolumn{3}{|c||}{Final state}\\
		\hline
		~2 & 6 & ~3 & 7 & ~5 & 9 & ~8 & 12 & Alice & Bob & \multicolumn{1}{|c||}{Form} \\
		\hline
		\hline
		\multicolumn{2}{|c|}{$\Phi^+$} & \multicolumn{2}{|c|}{$D$} & \multicolumn{2}{|c|}{$\Phi^+$} & \multicolumn{2}{|c||}{$D$} &\multicolumn{1}{|c|}{$1$} & \multicolumn{1}{|c|}{$10$} & \multicolumn{1}{|c||}{$\Phi^+$}\\
		\hline
		\multicolumn{2}{|c|}{$\Phi^+$} & \multicolumn{2}{|c|}{$D$} & \multicolumn{2}{|c|}{$\Phi^+$} & \multicolumn{2}{|c||}{$A$} & \multicolumn{1}{|c|}{$1$} & \multicolumn{1}{|c|}{$10$} & \multicolumn{1}{|c||}{$\Phi^-$}\\
		\hline
		\multicolumn{2}{|c|}{$\Phi^+$} & \multicolumn{2}{|c|}{$A$} & \multicolumn{2}{|c|}{$\Phi^+$} & \multicolumn{2}{|c||}{$D$} & \multicolumn{1}{|c|}{$1$} & \multicolumn{1}{|c|}{$10$} & \multicolumn{1}{|c||}{$\Phi^-$}\\
		\hline
		\multicolumn{2}{|c|}{$\Phi^+$} & \multicolumn{2}{|c|}{$A$} & \multicolumn{2}{|c|}{$\Phi^+$} & \multicolumn{2}{|c||}{$A$} & \multicolumn{1}{|c|}{$1$} & \multicolumn{1}{|c|}{$10$} & \multicolumn{1}{|c||}{$\Phi^+$}\\
		\hline
		\multicolumn{2}{|c|}{$\Phi^+$} & \multicolumn{2}{|c|}{$D$} & \multicolumn{2}{|c|}{$\Psi^+$} & \multicolumn{2}{|c||}{$D$} &\multicolumn{1}{|c|}{$1$} & \multicolumn{1}{|c|}{$10$} & \multicolumn{1}{|c||}{$\Psi^+$}\\
		\hline
		\multicolumn{2}{|c|}{$\Phi^+$} & \multicolumn{2}{|c|}{$D$} & \multicolumn{2}{|c|}{$\Psi^+$} & \multicolumn{2}{|c||}{$A$} &\multicolumn{1}{|c|}{$1$} & \multicolumn{1}{|c|}{$10$} & \multicolumn{1}{|c||}{$\Psi^-$}\\
		\hline
		\multicolumn{2}{|c|}{$\Phi^+$} & \multicolumn{2}{|c|}{$A$} & \multicolumn{2}{|c|}{$\Psi^+$} & \multicolumn{2}{|c||}{$D$} & \multicolumn{1}{|c|}{$1$} & \multicolumn{1}{|c|}{$10$} & \multicolumn{1}{|c||}{$\Psi^-$}\\
		\hline
		\multicolumn{2}{|c|}{$\Phi^+$} & \multicolumn{2}{|c|}{$A$} & \multicolumn{2}{|c|}{$\Psi^+$} & \multicolumn{2}{|c||}{$A$} & \multicolumn{1}{|c|}{$1$} & \multicolumn{1}{|c|}{$10$} & \multicolumn{1}{|c||}{$\Psi^+$}\\
		\hline
		\multicolumn{2}{|c|}{$\Psi^+$} & \multicolumn{2}{|c|}{$D$} & \multicolumn{2}{|c|}{$\Phi^+$} & \multicolumn{2}{|c||}{$D$} & \multicolumn{1}{|c|}{$1$} & \multicolumn{1}{|c|}{$10$} & \multicolumn{1}{|c||}{$\Psi^+$}\\
		\hline
		\multicolumn{2}{|c|}{$\Psi^+$} & \multicolumn{2}{|c|}{$D$} & \multicolumn{2}{|c|}{$\Phi^+$} & \multicolumn{2}{|c||}{$A$} & \multicolumn{1}{|c|}{$1$} & \multicolumn{1}{|c|}{$10$} & \multicolumn{1}{|c||}{$\Psi^-$}\\
		\hline
		\multicolumn{2}{|c|}{$\Psi^+$} & \multicolumn{2}{|c|}{$A$} & \multicolumn{2}{|c|}{$\Phi^+$} & \multicolumn{2}{|c||}{$D$} &\multicolumn{1}{|c|}{$1$} & \multicolumn{1}{|c|}{$10$} & \multicolumn{1}{|c||}{$\Psi^-$}\\
		\hline
		\multicolumn{2}{|c|}{$\Psi^+$} & \multicolumn{2}{|c|}{$A$} & \multicolumn{2}{|c|}{$\Phi^+$} & \multicolumn{2}{|c||}{$A$} &\multicolumn{1}{|c|}{$1$} & \multicolumn{1}{|c|}{$10$} & \multicolumn{1}{|c||}{$\Psi^+$}\\
		\hline
		\multicolumn{2}{|c|}{$\Psi^+$} & \multicolumn{2}{|c|}{$D$} & \multicolumn{2}{|c|}{$\Psi^+$} & \multicolumn{2}{|c||}{$D$} & \multicolumn{1}{|c|}{$1$} & \multicolumn{1}{|c|}{$10$} & \multicolumn{1}{|c||}{$\Phi^+$}\\
		\hline
		\multicolumn{2}{|c|}{$\Psi^+$} & \multicolumn{2}{|c|}{$D$} & \multicolumn{2}{|c|}{$\Psi^+$} & \multicolumn{2}{|c||}{$A$} & \multicolumn{1}{|c|}{$1$} & \multicolumn{1}{|c|}{$10$} & \multicolumn{1}{|c||}{$\Phi^-$}\\
		\hline
		\multicolumn{2}{|c|}{$\Psi^+$} & \multicolumn{2}{|c|}{$A$} & \multicolumn{2}{|c|}{$\Psi^+$} & \multicolumn{2}{|c||}{$D$} &\multicolumn{1}{|c|}{$1$} & \multicolumn{1}{|c|}{$10$} & \multicolumn{1}{|c||}{$\Phi^-$}\\
		\hline
		\multicolumn{2}{|c|}{$\Psi^+$} & \multicolumn{2}{|c|}{$A$} & \multicolumn{2}{|c|}{$\Psi^+$} & \multicolumn{2}{|c||}{$A$} & \multicolumn{1}{|c|}{$1$} & \multicolumn{1}{|c|}{$10$} & \multicolumn{1}{|c||}{$\Phi^+$}\\
		\hline
		\hline
		\multicolumn{2}{|c|}{$\Phi^+$} & \multicolumn{2}{|c|}{$D$} & \multicolumn{2}{|c|}{$D$} & \multicolumn{2}{|c||}{$\Phi^+$} & \multicolumn{1}{|c|}{$1$} & \multicolumn{1}{|c|}{$11$} & \multicolumn{1}{|c||}{$\Phi^+$}\\
		\hline
		\multicolumn{2}{|c|}{$\Phi^+$} & \multicolumn{2}{|c|}{$D$} & \multicolumn{2}{|c|}{$A$} & \multicolumn{2}{|c||}{$\Phi^+$} & \multicolumn{1}{|c|}{$1$} & \multicolumn{1}{|c|}{$11$} & \multicolumn{1}{|c||}{$\Phi^-$}\\
		\hline
		\multicolumn{2}{|c|}{$\Phi^+$} & \multicolumn{2}{|c|}{$A$} & \multicolumn{2}{|c|}{$D$} & \multicolumn{2}{|c||}{$\Phi^+$} &\multicolumn{1}{|c|}{$1$} & \multicolumn{1}{|c|}{$11$} & \multicolumn{1}{|c||}{$\Phi^-$}\\
		\hline
		\multicolumn{2}{|c|}{$\Phi^+$} & \multicolumn{2}{|c|}{$A$} & \multicolumn{2}{|c|}{$A$} & \multicolumn{2}{|c||}{$\Phi^+$} & \multicolumn{1}{|c|}{$1$} & \multicolumn{1}{|c|}{$11$} & \multicolumn{1}{|c||}{$\Phi^+$}\\
		\hline
		\multicolumn{2}{|c|}{$\Phi^+$} & \multicolumn{2}{|c|}{$D$} & \multicolumn{2}{|c|}{$D$} & \multicolumn{2}{|c||}{$\Psi^+$} & \multicolumn{1}{|c|}{$1$} & \multicolumn{1}{|c|}{$11$} & \multicolumn{1}{|c||}{$\Psi^+$}\\
		\hline
		\multicolumn{2}{|c|}{$\Phi^+$} & \multicolumn{2}{|c|}{$D$} & \multicolumn{2}{|c|}{$A$} & \multicolumn{2}{|c||}{$\Psi^+$} &\multicolumn{1}{|c|}{$1$} & \multicolumn{1}{|c|}{$11$} & \multicolumn{1}{|c||}{$\Psi^-$}\\
		\hline
		\multicolumn{2}{|c|}{$\Phi^+$} & \multicolumn{2}{|c|}{$A$} & \multicolumn{2}{|c|}{$D$} & \multicolumn{2}{|c||}{$\Psi^+$} &\multicolumn{1}{|c|}{$1$} & \multicolumn{1}{|c|}{$11$} & \multicolumn{1}{|c||}{$\Psi^-$}\\
		\hline
		\multicolumn{2}{|c|}{$\Phi^+$} & \multicolumn{2}{|c|}{$A$} & \multicolumn{2}{|c|}{$A$} & \multicolumn{2}{|c||}{$\Psi^+$} & \multicolumn{1}{|c|}{$1$} & \multicolumn{1}{|c|}{$11$} & \multicolumn{1}{|c||}{$\Psi^+$}\\
		\hline
		\multicolumn{2}{|c|}{$\Psi^+$} & \multicolumn{2}{|c|}{$D$} & \multicolumn{2}{|c|}{$D$} & \multicolumn{2}{|c||}{$\Phi^+$} & \multicolumn{1}{|c|}{$1$} & \multicolumn{1}{|c|}{$11$} & \multicolumn{1}{|c||}{$\Psi^+$}\\
		\hline
		\multicolumn{2}{|c|}{$\Psi^+$} & \multicolumn{2}{|c|}{$D$} & \multicolumn{2}{|c|}{$A$} & \multicolumn{2}{|c||}{$\Phi^+$} &\multicolumn{1}{|c|}{$1$} & \multicolumn{1}{|c|}{$11$} & \multicolumn{1}{|c||}{$\Psi^-$}\\
		\hline
		\multicolumn{2}{|c|}{$\Psi^+$} & \multicolumn{2}{|c|}{$A$} & \multicolumn{2}{|c|}{$D$} & \multicolumn{2}{|c||}{$\Phi^+$} &\multicolumn{1}{|c|}{$1$} & \multicolumn{1}{|c|}{$11$} & \multicolumn{1}{|c||}{$\Psi^-$}\\
		\hline
		\multicolumn{2}{|c|}{$\Psi^+$} & \multicolumn{2}{|c|}{$A$} & \multicolumn{2}{|c|}{$A$} & \multicolumn{2}{|c||}{$\Phi^+$} & \multicolumn{1}{|c|}{$1$} & \multicolumn{1}{|c|}{$11$} & \multicolumn{1}{|c||}{$\Psi^+$}\\
		\hline
		\multicolumn{2}{|c|}{$\Psi^+$} & \multicolumn{2}{|c|}{$D$} & \multicolumn{2}{|c|}{$D$} & \multicolumn{2}{|c||}{$\Psi^+$} & \multicolumn{1}{|c|}{$1$} & \multicolumn{1}{|c|}{$11$} & \multicolumn{1}{|c||}{$\Phi^+$}\\
		\hline
		\multicolumn{2}{|c|}{$\Psi^+$} & \multicolumn{2}{|c|}{$D$} & \multicolumn{2}{|c|}{$A$} & \multicolumn{2}{|c||}{$\Psi^+$} &\multicolumn{1}{|c|}{$1$} & \multicolumn{1}{|c|}{$11$} & \multicolumn{1}{|c||}{$\Phi^-$}\\
		\hline
		\multicolumn{2}{|c|}{$\Psi^+$} & \multicolumn{2}{|c|}{$A$} & \multicolumn{2}{|c|}{$D$} & \multicolumn{2}{|c||}{$\Psi^+$} &\multicolumn{1}{|c|}{$1$} & \multicolumn{1}{|c|}{$11$} & \multicolumn{1}{|c||}{$\Phi^-$}\\
		\hline
		\multicolumn{2}{|c|}{$\Psi^+$} & \multicolumn{2}{|c|}{$A$} & \multicolumn{2}{|c|}{$A$} & \multicolumn{2}{|c||}{$\Psi^+$} &\multicolumn{1}{|c|}{$1$} & \multicolumn{1}{|c|}{$11$} & \multicolumn{1}{|c||}{$\Phi^+$}\\
		\hline

		\multicolumn{2}{|c|}{$D$} & \multicolumn{2}{|c|}{$\Phi^+$} & \multicolumn{2}{|c|}{$\Phi^+$} & \multicolumn{2}{|c||}{$D$} &\multicolumn{1}{|c|}{$4$} & \multicolumn{1}{|c|}{$10$} & \multicolumn{1}{|c||}{$\Phi^+$}\\
		\hline
		\multicolumn{2}{|c|}{$D$} & \multicolumn{2}{|c|}{$\Phi^+$} & \multicolumn{2}{|c|}{$\Phi^+$} & \multicolumn{2}{|c||}{$A$} &\multicolumn{1}{|c|}{$4$} & \multicolumn{1}{|c|}{$10$} & \multicolumn{1}{|c||}{$\Phi^-$}\\
		\hline
		\multicolumn{2}{|c|}{$A$} & \multicolumn{2}{|c|}{$\Phi^+$} & \multicolumn{2}{|c|}{$\Phi^+$} & \multicolumn{2}{|c||}{$D$} &\multicolumn{1}{|c|}{$4$} & \multicolumn{1}{|c|}{$10$} & \multicolumn{1}{|c||}{$\Phi^-$}\\
		\hline
		\multicolumn{2}{|c|}{$A$} & \multicolumn{2}{|c|}{$\Phi^+$} & \multicolumn{2}{|c|}{$\Phi^+$} & \multicolumn{2}{|c||}{$A$} &\multicolumn{1}{|c|}{$4$} & \multicolumn{1}{|c|}{$10$} & \multicolumn{1}{|c||}{$\Phi^+$}\\
		\hline
		
		\multicolumn{2}{|c|}{$D$} & \multicolumn{2}{|c|}{$\Phi^+$} & \multicolumn{2}{|c|}{$\Psi^+$} & \multicolumn{2}{|c||}{$D$} &\multicolumn{1}{|c|}{$4$} & \multicolumn{1}{|c|}{$10$} & \multicolumn{1}{|c||}{$\Psi^+$}\\
		\hline
		\multicolumn{2}{|c|}{$D$} & \multicolumn{2}{|c|}{$\Phi^+$} & \multicolumn{2}{|c|}{$\Psi^+$} & \multicolumn{2}{|c||}{$A$} &\multicolumn{1}{|c|}{$4$} & \multicolumn{1}{|c|}{$10$} & \multicolumn{1}{|c||}{$\Psi^-$}\\
		\hline
		\multicolumn{2}{|c|}{$A$} & \multicolumn{2}{|c|}{$\Phi^+$} & \multicolumn{2}{|c|}{$\Psi^+$} & \multicolumn{2}{|c||}{$D$} &\multicolumn{1}{|c|}{$4$} & \multicolumn{1}{|c|}{$10$} & \multicolumn{1}{|c||}{$\Psi^-$}\\
		\hline
		\multicolumn{2}{|c|}{$A$} & \multicolumn{2}{|c|}{$\Phi^+$} & \multicolumn{2}{|c|}{$\Psi^+$} & \multicolumn{2}{|c||}{$A$} &\multicolumn{1}{|c|}{$4$} & \multicolumn{1}{|c|}{$10$} & \multicolumn{1}{|c||}{$\Psi^+$}\\
		\hline
		
		\multicolumn{2}{|c|}{$D$} & \multicolumn{2}{|c|}{$\Psi^+$} & \multicolumn{2}{|c|}{$\Phi^+$} & \multicolumn{2}{|c||}{$D$} &\multicolumn{1}{|c|}{$4$} & \multicolumn{1}{|c|}{$10$} & \multicolumn{1}{|c||}{$\Psi^+$}\\
		\hline
		\multicolumn{2}{|c|}{$D$} & \multicolumn{2}{|c|}{$\Psi^+$} & \multicolumn{2}{|c|}{$\Phi^+$} & \multicolumn{2}{|c||}{$A$} &\multicolumn{1}{|c|}{$4$} & \multicolumn{1}{|c|}{$10$} & \multicolumn{1}{|c||}{$\Psi^-$}\\
		\hline
		\multicolumn{2}{|c|}{$A$} & \multicolumn{2}{|c|}{$\Psi^+$} & \multicolumn{2}{|c|}{$\Phi^+$} & \multicolumn{2}{|c||}{$D$} &\multicolumn{1}{|c|}{$4$} & \multicolumn{1}{|c|}{$10$} & \multicolumn{1}{|c||}{$\Psi^-$}\\
		\hline
		\multicolumn{2}{|c|}{$A$} & \multicolumn{2}{|c|}{$\Psi^+$} & \multicolumn{2}{|c|}{$\Phi^+$} & \multicolumn{2}{|c||}{$A$} &\multicolumn{1}{|c|}{$4$} & \multicolumn{1}{|c|}{$10$} & \multicolumn{1}{|c||}{$\Psi^+$}\\
		\hline
		
		\multicolumn{2}{|c|}{$D$} & \multicolumn{2}{|c|}{$\Psi^+$} & \multicolumn{2}{|c|}{$\Psi^+$} & \multicolumn{2}{|c||}{$D$} &\multicolumn{1}{|c|}{$4$} & \multicolumn{1}{|c|}{$10$} & \multicolumn{1}{|c||}{$\Phi^+$}\\
		\hline
		\multicolumn{2}{|c|}{$D$} & \multicolumn{2}{|c|}{$\Psi^+$} & \multicolumn{2}{|c|}{$\Psi^+$} & \multicolumn{2}{|c||}{$A$} &\multicolumn{1}{|c|}{$4$} & \multicolumn{1}{|c|}{$10$} & \multicolumn{1}{|c||}{$\Phi^-$}\\
		\hline
		\multicolumn{2}{|c|}{$A$} & \multicolumn{2}{|c|}{$\Psi^+$} & \multicolumn{2}{|c|}{$\Psi^+$} & \multicolumn{2}{|c||}{$D$} &\multicolumn{1}{|c|}{$4$} & \multicolumn{1}{|c|}{$10$} & \multicolumn{1}{|c||}{$\Phi^-$}\\
		\hline
		\multicolumn{2}{|c|}{$A$} & \multicolumn{2}{|c|}{$\Psi^+$} & \multicolumn{2}{|c|}{$\Psi^+$} & \multicolumn{2}{|c||}{$A$} &\multicolumn{1}{|c|}{$4$} & \multicolumn{1}{|c|}{$10$} & \multicolumn{1}{|c||}{$\Phi^+$}\\
		\hline

		\multicolumn{2}{|c|}{$D$} & \multicolumn{2}{|c|}{$\Phi^+$} & \multicolumn{2}{|c|}{$D$} & \multicolumn{2}{|c||}{$\Phi^+$} &\multicolumn{1}{|c|}{$4$} & \multicolumn{1}{|c|}{$11$} & \multicolumn{1}{|c||}{$\Phi^+$}\\
		\hline
		\multicolumn{2}{|c|}{$D$} & \multicolumn{2}{|c|}{$\Phi^+$} & \multicolumn{2}{|c|}{$A$} & \multicolumn{2}{|c||}{$\Phi^+$} &\multicolumn{1}{|c|}{$4$} & \multicolumn{1}{|c|}{$11$} & \multicolumn{1}{|c||}{$\Phi^-$}\\
		\hline
		\multicolumn{2}{|c|}{$A$} & \multicolumn{2}{|c|}{$\Phi^+$} & \multicolumn{2}{|c|}{$D$} & \multicolumn{2}{|c||}{$\Phi^+$} &\multicolumn{1}{|c|}{$4$} & \multicolumn{1}{|c|}{$11$} & \multicolumn{1}{|c||}{$\Phi^-$}\\
		\hline
		\multicolumn{2}{|c|}{$A$} & \multicolumn{2}{|c|}{$\Phi^+$} & \multicolumn{2}{|c|}{$A$} & \multicolumn{2}{|c||}{$\Phi^+$} &\multicolumn{1}{|c|}{$4$} & \multicolumn{1}{|c|}{$11$} & \multicolumn{1}{|c||}{$\Phi^+$}\\
		\hline
		
		\multicolumn{2}{|c|}{$D$} & \multicolumn{2}{|c|}{$\Phi^+$} & \multicolumn{2}{|c|}{$D$} & \multicolumn{2}{|c||}{$\Psi^+$} &\multicolumn{1}{|c|}{$4$} & \multicolumn{1}{|c|}{$11$} & \multicolumn{1}{|c||}{$\Psi^+$}\\
		\hline
		\multicolumn{2}{|c|}{$D$} & \multicolumn{2}{|c|}{$\Phi^+$} & \multicolumn{2}{|c|}{$A$} & \multicolumn{2}{|c||}{$\Psi^+$} &\multicolumn{1}{|c|}{$4$} & \multicolumn{1}{|c|}{$11$} & \multicolumn{1}{|c||}{$\Psi^-$}\\
		\hline
		\multicolumn{2}{|c|}{$A$} & \multicolumn{2}{|c|}{$\Phi^+$} & \multicolumn{2}{|c|}{$D$} & \multicolumn{2}{|c||}{$\Psi^+$} &\multicolumn{1}{|c|}{$4$} & \multicolumn{1}{|c|}{$11$} & \multicolumn{1}{|c||}{$\Psi^-$}\\
		\hline
		\multicolumn{2}{|c|}{$A$} & \multicolumn{2}{|c|}{$\Phi^+$} & \multicolumn{2}{|c|}{$A$} & \multicolumn{2}{|c||}{$\Psi^+$} &\multicolumn{1}{|c|}{$4$} & \multicolumn{1}{|c|}{$11$} & \multicolumn{1}{|c||}{$\Psi^+$}\\
		\hline
		
		\multicolumn{2}{|c|}{$D$} & \multicolumn{2}{|c|}{$\Psi^+$} & \multicolumn{2}{|c|}{$D$} & \multicolumn{2}{|c||}{$\Phi^+$} &\multicolumn{1}{|c|}{$4$} & \multicolumn{1}{|c|}{$11$} & \multicolumn{1}{|c||}{$\Psi^+$}\\
		\hline
		\multicolumn{2}{|c|}{$D$} & \multicolumn{2}{|c|}{$\Psi^+$} & \multicolumn{2}{|c|}{$A$} & \multicolumn{2}{|c||}{$\Phi^+$} &\multicolumn{1}{|c|}{$4$} & \multicolumn{1}{|c|}{$11$} & \multicolumn{1}{|c||}{$\Psi^-$}\\
		\hline
		\multicolumn{2}{|c|}{$A$} & \multicolumn{2}{|c|}{$\Psi^+$} & \multicolumn{2}{|c|}{$D$} & \multicolumn{2}{|c||}{$\Phi^+$} &\multicolumn{1}{|c|}{$4$} & \multicolumn{1}{|c|}{$11$} & \multicolumn{1}{|c||}{$\Psi^-$}\\
		\hline
		\multicolumn{2}{|c|}{$A$} & \multicolumn{2}{|c|}{$\Psi^+$} & \multicolumn{2}{|c|}{$A$} & \multicolumn{2}{|c||}{$\Phi^+$} &\multicolumn{1}{|c|}{$4$} & \multicolumn{1}{|c|}{$11$} & \multicolumn{1}{|c||}{$\Psi^+$}\\
		\hline
		
		\multicolumn{2}{|c|}{$D$} & \multicolumn{2}{|c|}{$\Psi^+$} & \multicolumn{2}{|c|}{$D$} & \multicolumn{2}{|c||}{$\Psi^+$} &\multicolumn{1}{|c|}{$4$} & \multicolumn{1}{|c|}{$11$} & \multicolumn{1}{|c||}{$\Phi^+$}\\
		\hline
		\multicolumn{2}{|c|}{$D$} & \multicolumn{2}{|c|}{$\Psi^+$} & \multicolumn{2}{|c|}{$A$} & \multicolumn{2}{|c||}{$\Psi^+$} &\multicolumn{1}{|c|}{$4$} & \multicolumn{1}{|c|}{$11$} & \multicolumn{1}{|c||}{$\Phi^-$}\\
		\hline
		\multicolumn{2}{|c|}{$A$} & \multicolumn{2}{|c|}{$\Psi^+$} & \multicolumn{2}{|c|}{$D$} & \multicolumn{2}{|c||}{$\Psi^+$} &\multicolumn{1}{|c|}{$4$} & \multicolumn{1}{|c|}{$11$} & \multicolumn{1}{|c||}{$\Phi^-$}\\
		\hline
		\multicolumn{2}{|c|}{$A$} & \multicolumn{2}{|c|}{$\Psi^+$} & \multicolumn{2}{|c|}{$A$} & \multicolumn{2}{|c||}{$\Psi^+$} & $4$ & \multicolumn{1}{|c|}{$11$} & \multicolumn{1}{|c||}{$\Phi^+$}\\
		\hline
\end{longtable}}

\section{The signal-to-noise ratio of 12 photons}

\begin{figure*}[htbp]
	\centering
	\includegraphics[width=0.9\linewidth]{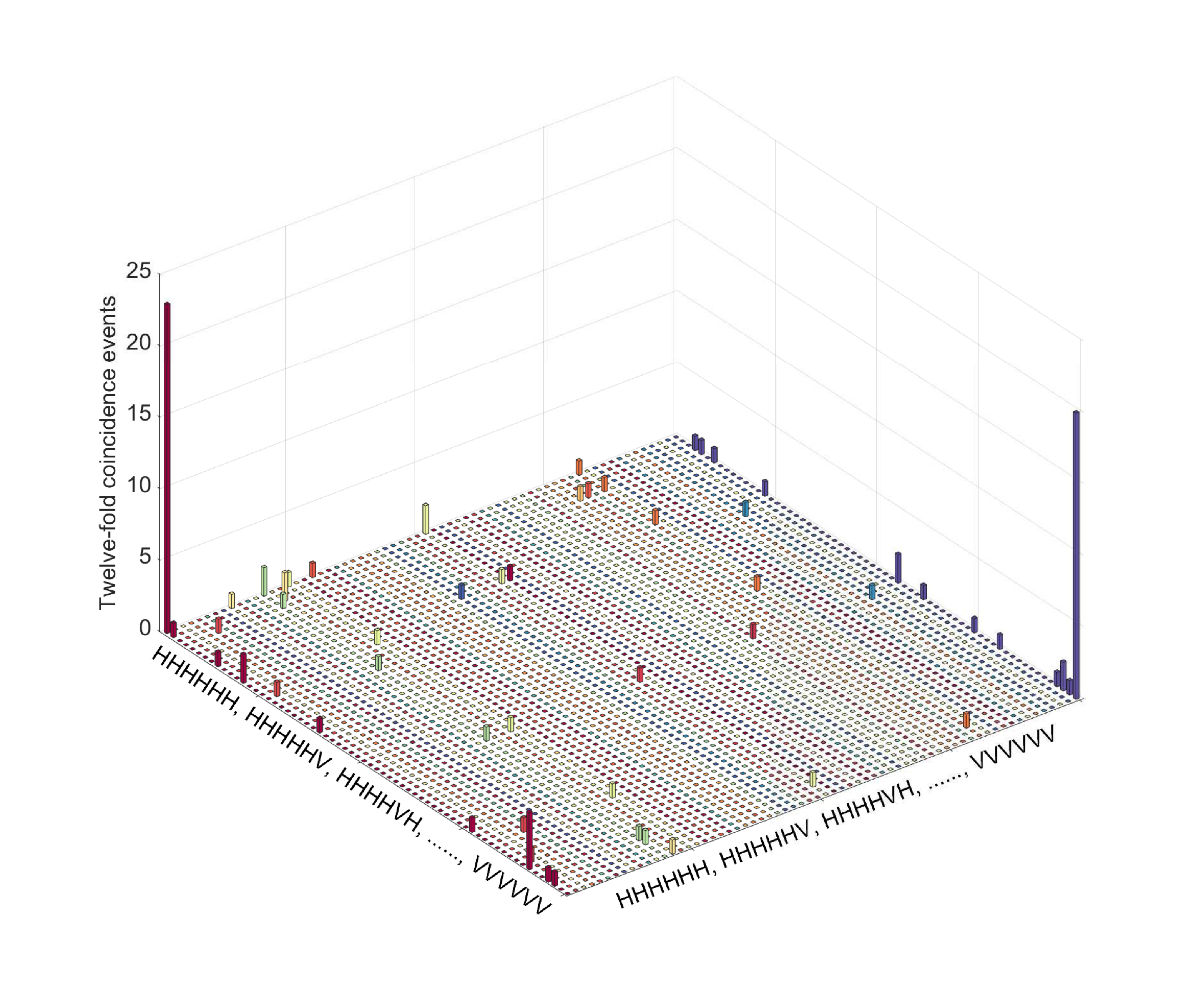}
	\caption{\textbf{12-photon coincidence counts measured in the $\ket{H}/\ket{V}$ basis}}
	\label{fig:12photon}
\end{figure*}

In our experiment, the whole setup can be considered as a twelve-photon interferometer. To obtain a higher coincidence rate, just bandpass filters with $\lambda_{\text{FWHM}} = 30~\text{nm}$ are set before the single photon detector to remove the small sidebands. With a pump power of 700 mW we obtain a 12-photon coincidence rate of 1.2 $\text{h}^{-1}$.
To verify the ability of experimentally manipulating twelve photons, we turn the circular PBS to a normal PBS by setting the surrounding HWPs at $0\degree$. Then the whole setup serves for generating twelve-photon mix states
\begin{equation}
\begin{aligned}
\rho_{12}=v(\ket{H^{\otimes12}}+\ket{V^{\otimes12}})(\bra{H^{\otimes12}}+\bra{V^{\otimes12}})+\frac{(1-v)}{2}(\ket{H^{\otimes12}}\bra{H^{\otimes12}}+\ket{V^{\otimes12}}\bra{V^{\otimes12}}).
\end{aligned}
\end{equation}
To characterize this state, we measure it in $Z^{\otimes12}$ basis and show that under the condition of registering a 12-fold coincidence only the $\ket{H^{\otimes12}}$ and $\ket{V^{\otimes12}}$ components can be observed, but no others. The experimental data are shown in Fig.~\ref{fig:12photon}. The signal-to-noise ratio defined as the ratio of the average of the desired components to that of the non-desired ones is $1420:1$. It indicates that the twelve photons can be well manipulated in our set-up.

\section{Detail results with down-conversion probability $p=0.0483$}
\begin{figure*}[htbp]
	\centering
	\includegraphics[width=0.8\linewidth]{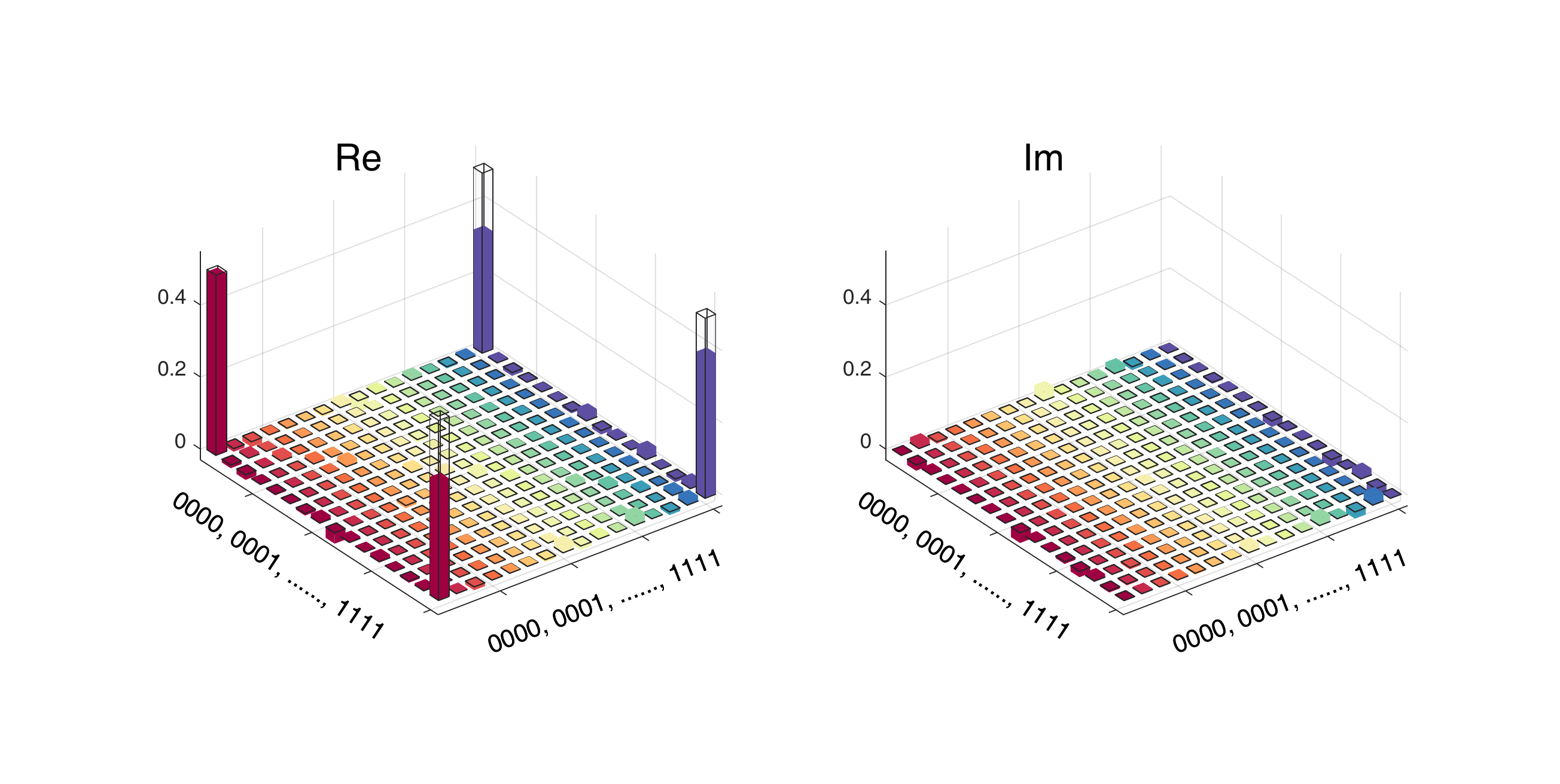}
	\caption{\textbf{Reconstructed density matrix of 4-photon GHZ state with $p=0.0483\pm0.0001$.}}
	\label{fig:GHZ-tomo-720}
\end{figure*}

\begin{figure*}[htbp]
	\centering
	\includegraphics[width=0.8\linewidth]{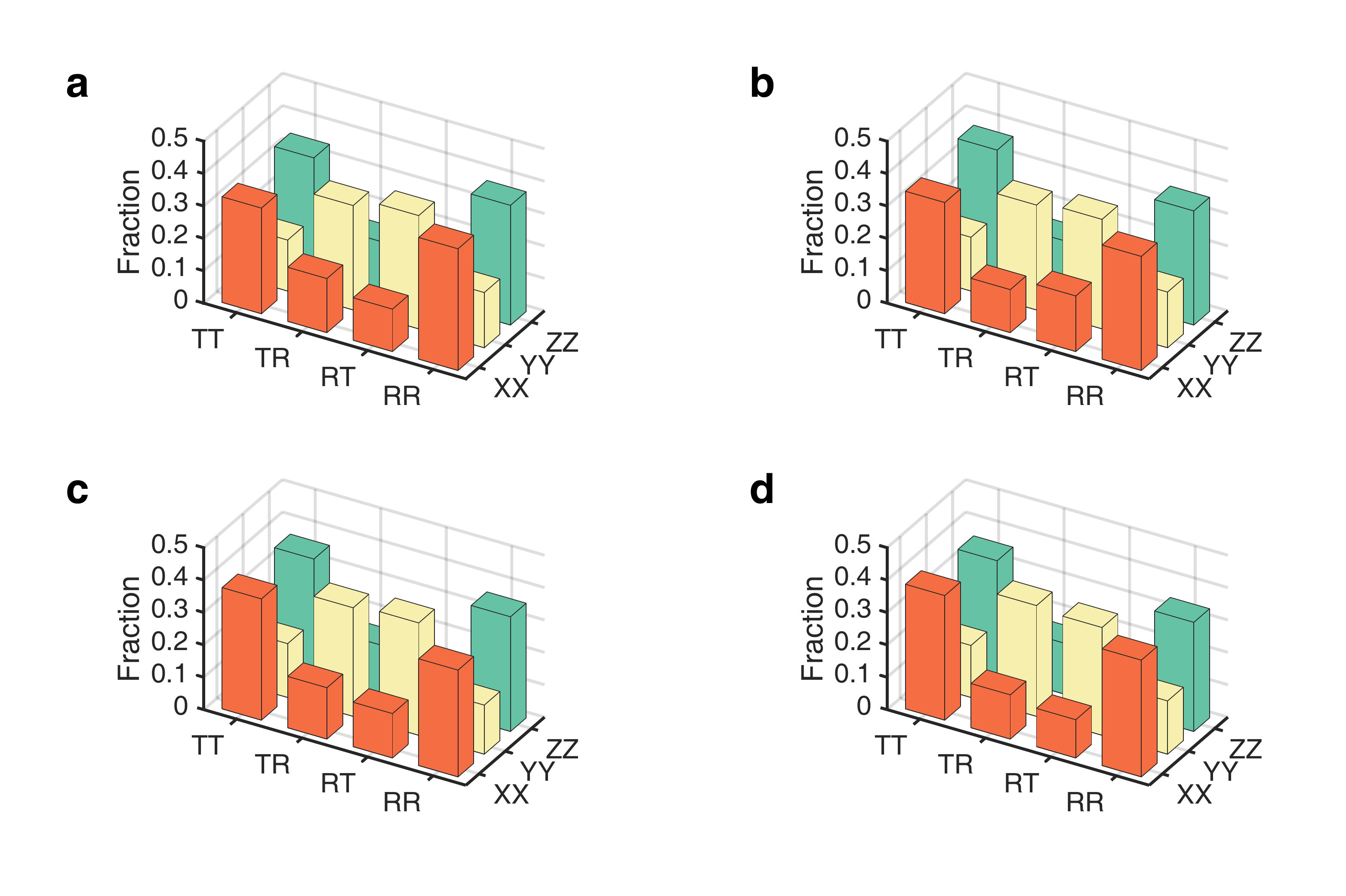}
	\caption{\textbf{The measured fractions of final entangled photon pair with $p=0.049$.}
		\textbf{a-d,} The measured fractions in basis XX, YY and ZZ with final entangled photon pairs 1\&11, 4\&11, 1\&10 and 4\&10, respectively. {The corresponding fidelities are $0.542\pm0.012$, $0.537\pm0.010$, $0.549\pm0.012$ and $0.554\pm0.010$.}}
	\label{fig:result-720}
\end{figure*}

Due to the higher-order noise of SPDC process, the ratio $r$ of entanglement generation rate between all-photonic quantum repeater and conventional scheme is also influenced by the down-conversion probability, $p$.
To verify the relationship of $r$ and $p$, we increase the power of pump laser to $720~mW$ and acquire another data point with $p=0.049$ (shown in Fig. \ref{fig:result}a in main text).
For a detailed characterization of the $2\times2$ all-photonic repeater with $p=0.049$, we also do tomographic measurements~\cite{James2001PRA} on the four-photon GHZ state. The reconstructed matrix is shown in Fig. \ref{fig:GHZ-tomo-720} and the fidelity is calculated as $F=0.877$, which indicates the 4-photon GHZ states are genuine entangled.
At last, we also examine the entanglement of final photon pairs. In experiment We register $22~\text{h}$ eight-fold coincidence in XX, YY and ZZ basis respectively. The measured fraction of final Entangled photon pairs 1\&11, 4\&11, 1\&10, 4\&10 are shown in Fig. \ref{fig:result-720}a-d and the fidelity is calculated as $0.546\pm0.006$. It indicates the $2\times2$ parallel structured all-photonic quantum repeater is fully demonstrated with the down-conversion probability$p=0.0483\pm0.0001$.

%

%

\end{document}